\documentclass[aps,prd,10pt,nofootinbib,twocolumn,superscriptaddress,floatfix,notitlepage]{revtex4-1}

\usepackage{graphicx}
\usepackage{amsmath,amsfonts,amssymb}
\usepackage{xcolor}
\usepackage[breaklinks,colorlinks,urlcolor=blue,citecolor=blue,linkcolor=magenta]{hyperref}
\usepackage{verbatim}
\usepackage{aas_macros}

\newcommand{\td}{{\rm d}}
\newcommand{\vect}[1]{\boldsymbol{#1}}

\newcommand{\be}{\begin{equation}}
\newcommand{\ee}{\end{equation}}
\newcommand{\bea}{\begin{equation} \begin{aligned}}
\newcommand{\eea}{\end{aligned} \end{equation}}

\newcommand{\X}{\vect{X}}             
\newcommand{\q}{\vect{q}}

\newcommand{\kvec}{\vect{k}}
\newcommand{\Ptil}{\tilde P}

\begin{document}

\title{Cosmological lensing as a jump process}

\author{Galymzhan Baltabay}
\email{galymzhan.baltabay@kbfi.ee}
\affiliation{Laboratory of High Energy and Computational Physics, KBFI, R{\"a}vala 10, Tallinn, 10143, Estonia}
\affiliation{Department of Cybernetics, Tallinn University of Technology, Akadeemia tee 21, 12618 Tallinn, Estonia}

\author{Ville Vaskonen}
\email{ville.vaskonen@kbfi.ee}
\affiliation{Laboratory of High Energy and Computational Physics, KBFI, R{\"a}vala 10, Tallinn, 10143, Estonia}

\begin{abstract}
We develop a jump-process formalism for the probability density function (PDF) of gravitational lensing fields. In the Born approximation, each lens adds a jump to the convergence $\kappa$ and the two components of the shear, making lensing a jump process in redshift. For independent lenses, the joint PDF of the lensing fields obeys a compensated Kolmogorov-Feller equation, which we solve in closed form in Fourier space and extend to clustered lenses using their two-point correlation function. A Fourier-Hankel inversion then yields the joint PDF of $\kappa$ and the shear magnitude $\gamma$, from which the magnification PDF follows, including its high-magnification tail, without sampling noise. We use these PDFs to train an updated version of the \texttt{FLUMEN} emulator, which makes likelihood-based inference from the full magnification distribution feasible across cosmologies. To motivate such analyses, we estimate that samples of $10^4$ Roman supernovae, $10^5$ Rubin supernovae, 300 Einstein Telescope standard sirens, and 12 LISA standard sirens could constrain the matter fluctuation amplitude $\sigma_8$ respectively to 4.8\%, 5.1\%, 8.5\%, and 19\%, with the other cosmological parameters fixed.
\end{abstract}

\maketitle

% ============================================================================
\section{Introduction}
\label{sec:intro}
% ============================================================================
Gravitational lensing by intervening structure affects the signals from distant sources. The mean effect is set by the average matter distribution, while the scatter around it carries information on the abundance and clustering of structure. The magnification PDF is therefore used as a probe with supernovae~\cite{Hamana:1999rk,Wang:2004ax,Dodelson:2005zt,Quartin:2013moa,Castro:2014oja,Zhai:2019sgi,DES:2024lto}, quasars~\cite{Hamana:1999bf}, and gravitational wave (GW) sources~\cite{Congedo:2018wfn,Mpetha:2024xiu,Vaskonen:2026ubd}. This requires an accurate estimate of the magnification PDF $\td P/\td\mu$, which is usually obtained by sampling: by ray tracing through $N$-body simulations~\cite{Holz:1997ic,Holz:2004xx,Hilbert:2007ny,Takahashi:2011qd,Takahashi:2017hjr}, or with the stochastic Monte Carlo approach of Refs.~\cite{Kainulainen:2009dw,Kainulainen:2010at,Kainulainen:2011zx,Vaskonen:2026ubd,Baltabay:2026nmz}. Each cosmology then needs a fresh set of realizations. Emulators~\cite{Turker:2025rdt,Baltabay:2026nmz} remove this cost, but inherit the noise and reach of their training data. Sampling-free predictions exist for parts of the problem: fitting functions~\cite{Wang:2002qc,Das:2005yb,Takahashi:2011qd}, fits to the moments~\cite{Marra:2013roi}, the one-point convergence PDF from hierarchical clustering~\cite{Valageas:1999ir,Valageas:1999ch,Munshi:1999qw}, the halo model~\cite{Thiele:2020rig}, large-deviation theory~\cite{Barthelemy:2019ciu,Boyle:2020bqn}, and, for point masses at low optical depth, the classical characteristic-function statistics~\cite{Nityananda:1984xx,Katz:1986xx,Schneider:1987xx,Kofman:1996es,Lee:1996et}.

In this work we compute $\td P/\td\mu$ without sampling. We treat the lenses as a marked point process along the line of sight, as already implicit in the stochastic approach~\cite{Kainulainen:2009dw}, which wrote the convergence PDF as a sum over Poisson occupation numbers. Each lens adds a jump to the convergence and shear, and their joint PDF obeys a master equation of the Kolmogorov-Feller type. It is the discrete-lens analogue of the Fokker-Planck equation for stochastic lensing in Ref.~\cite{Fleury:2015rwa}, which treats the lenses as Gaussian white noise, corresponding to the diffusion limit in which each lens contributes an infinitesimal distortion and only the mean and variance of the lens contributions are retained. In Fourier space it is solved in closed form for every wave vector. The lens population enters through the lens jumps and rates, and lens clustering adds a two-point term, which we evaluate with the linear power spectrum in the Limber approximation. We use the lens populations of Refs.~\cite{Vaskonen:2026ubd,Baltabay:2026nmz}, with truncated halos and filaments.

Compound-Poisson characteristic functions of lens populations are not new in themselves. They underlie the classical statistics of point-mass lenses at low optical depth~\cite{Nityananda:1984xx,Katz:1986xx,Schneider:1987xx,Kofman:1996es,Lee:1996et} and the convergence PDF of Ref.~\cite{Kainulainen:2009dw}, and in the halo model they give the one-point PDFs of the thermal Sunyaev-Zel'dovich signal~\cite{Thiele:2018jdl} and of the smoothed lensing convergence~\cite{Thiele:2020rig}, the latter including a linear-bias two-halo term. What is new here is that (i) we follow the full distortion vector, so that the magnification of a point source is built from its exact relation to the convergence and shear rather than from the convergence alone; (ii) the lenses include elliptical halos and cylindrical filaments; (iii) a host halo and its subhalos enter as one composite lens that conserves mass; (iv) the two-point clustering term keeps the transverse extent of the lenses and the spin structure of the shear; and (v) all of this follows from a single master equation in the lens redshift that keeps the jumps of the individual lenses whole, unlike the diffusion limit of Ref.~\cite{Fleury:2015rwa} which loses the skewness and the tail of the distribution. For a given lens model, the formalism yields the exact PDF, with no sampling and no further approximation beyond the truncation of the clustering expansion.

% ============================================================================
\section{Jump-process formalism}
\label{sec:formalism}
% ============================================================================

Lensing by structures near the line of sight changes the flux of an electromagnetic source, or the squared strain amplitude of a GW source, by a factor $\mu$ relative to a homogeneous universe. The magnification of a point source is set by the convergence $\kappa$ and the shear $(\gamma_1,\gamma_2)$ along its line of sight, which together form the Jacobian of the lens mapping. Its inverse determinant gives
\be \label{eq:mu}
    \mu = \frac{1}{|(1-\kappa)^2-\gamma^2|} \,, \qquad \gamma^2 = \gamma_1^2+\gamma_2^2 \,,
\ee
where the modulus makes $\mu$ positive for images of either parity.

In the Born approximation, where lensing fields are evaluated along the unperturbed line of sight, the convergence and shear contributions from individual lenses add linearly. We define
\be \label{eq:Xdef}
    \X \equiv \sum_j \q_j \,, \quad \q_j \equiv (\kappa_j,\gamma_{j,1},\gamma_{j,2}) \,,
\ee
where $j$ labels the lenses. We collect the lens impact parameter and profile parameters into $\theta_j$, such that $\q_j = \q(z_j,\theta_j;z_s)$, where $z_j$ denotes the lens redshift and $z_s$ the source redshift.

Assuming that lenses occur independently along the line of sight, a lens with parameters $\theta$ appears between $z$ and $z+\td z$ with probability $n(z,\theta)\td z\td\theta$. The cumulative lensing fields therefore evolve through independent stochastic jumps, giving the compensated Kolmogorov–Feller equation~\cite{Feller1940}
\bea \label{eq:kf}
    \partial_z P(\X;z) = & \int\!\td\theta\, n(z,\theta) \Big[ P(\X-\q;z) - P(\X;z) \\
    &\qquad\qquad\qquad\quad + \sum_a q_a\,\partial_a P(\X;z) \Big] ,
\eea
with $a \in \{\kappa,\gamma_1,\gamma_2\}$, $\partial_a = \partial/\partial X_a$. The first two terms describe the gain and loss of probability due to lensing jumps, while the third, the drift term, subtracts the mean contribution of the jumps, ensuring $\langle\X\rangle=0$. The initial condition is $P(\X;0) = \delta_{\rm D}(\X)$, corresponding to vanishing lensing fields at the observer. Unlike the Monte Carlo of Ref.~\cite{Vaskonen:2026ubd}, which follows only the lenses above a convergence threshold and adds the rest as a Gaussian background, Eq.~\eqref{eq:kf} needs no threshold.

In Fourier space with the characteristic function $\Ptil(\kvec;z) = \int\td^3X\,e^{i\kvec\cdot\X}P(\X;z)$, Eq.~\eqref{eq:kf} becomes one ordinary differential equation per wave vector,
\bea \label{eq:kf-k}
    \partial_z \Ptil(\kvec;z) &= \ell_1(\kvec;z)\,\Ptil(\kvec;z) \,, \\
    \ell_1(\kvec;z) &= \int\!\td\theta\, n(z,\theta)
      \Big[ e^{\,i\kvec\cdot\q} - 1 - i\kvec\cdot\q \Big] \,.
\eea
The solution at the source redshift $z_s$ is
\be \label{eq:kf-sol}
    \Ptil(\kvec;z_s) = e^{\Psi_1(\kvec)} \,, \quad \Psi_1(\kvec) = \int_0^{z_s}\!\td z\,\ell_1(\kvec;z) \,.
\ee
Taylor expanding the integrand of $\ell_1$ about $\kvec = 0$ shows that $\Psi_1$ generates all $n \ge 2$ cumulants of $\X$, $\langle X_{a_1}\cdots X_{a_n}\rangle_c = \int_0^{z_s}\td z\int\td\theta\, n\,q_{a_1}\cdots q_{a_n}$. Keeping only $n=2$ gives a Fokker-Planck equation and a Gaussian $P$, the diffusion limit of Ref.~\cite{Fleury:2015rwa} which misses the skewed shape of ${\rm d}P/{\rm d}\mu$. Truncating at any finite $n > 2$ is not an option either, as by Pawula's theorem~\cite{Pawula:1967zz} such truncations do not preserve the positivity of $P$. The jump term has to be kept whole.

When clustering of lenses is taken into account, the increments of $\X$ in successive redshift intervals are no longer independent and Eq.~\eqref{eq:kf} does not hold. The exponential form of the characteristic function $\Ptil$ survives since $\Ptil$ is a generating functional of the lens distribution. Its logarithm is, for any point process, a sum over connected correlation functions of the lens population,
\be \label{eq:psi-series}
    \ln\Ptil(\kvec;z_s) = \Psi_1(\kvec)+\Psi_2(\kvec) + \cdots \,,
\ee
where $\Psi_1$ is the Poisson term of Eq.~\eqref{eq:kf-sol} with $n$ now the mean lens density, and $\Psi_2$ involves the two-point correlation $\xi$ between lens pairs. All $\Psi_{n\ge2}$ vanish for Poisson lenses. We evaluate the two-point term $\Psi_2$ in Sec.~\ref{sec:clustering}.

% ============================================================================
\section{The lens population}
\label{sec:population}
% ============================================================================

We now apply the jump-process formalism to the lens model of Refs.~\cite{Vaskonen:2026ubd,Baltabay:2026nmz}: field halos with elliptical projections, filaments, subhalos, and the large-scale clustering of the lens population. Halos and filaments are two populations of jumps, subhalos turn each halo into a composite jump, and clustering correlates the lenses. We summarize the lens prescriptions of Refs.~\cite{Vaskonen:2026ubd,Baltabay:2026nmz} only where needed to specify these elements.

\subsection{Halos and filaments}
\label{sec:halos}

Field halos and filaments form two independent Poisson populations of lenses, $p = h, f$, correlated with each other only through the clustering of Sec.~\ref{sec:clustering}. A lens is described by its mass $M$, its position $\vect r = r (\cos\phi,\sin\phi)$ in the lens plane, measured from the line of sight, and an orientation angle $\varphi$. Halos are modeled by the pseudo-elliptical NFW profile~\cite{Navarro:1996gj,Golse:2001ar,Allgood:2005eu} truncated at $r_{200}$, and filaments, following the simulations of~\cite{Colberg:2004cd}, by cylinders with the radial density profile $\rho \propto (1 + r_\perp^2/r_s^2)^{-1}$, where $r_\perp$ is the distance from the filament axis and $r_s$ the scale radius, truncated at $r_\perp = r_s$. The expected number of lenses is\footnote{We use units with $c = G = 1$.}
\be \label{eq:dN}
    n_p(z,\theta)\,\td z\,\td\theta = \frac{(1+z)^2}{H(z)}
    \frac{\td\bar n_p}{\td \ln M}\;\td z\,\td \ln M \td^2 r \,\frac{\td\varphi}{2\pi} ,
\ee
where $\td\bar n_p/\td \ln M$ is the comoving mass function of the population~\cite{Sheth:1999mn,Shen:2005wd}. 

In both populations the shear is oriented by the polar angle $\phi$ of the lens position, so at fixed $\varphi$ the average over $\phi$ gives a Bessel function,
\be \label{eq:phiavg}
    \frac{1}{2\pi}\int_0^{2\pi}\!\td\phi\;
    e^{\,i(k_1\gamma_1 + k_2\gamma_2)} = J_0(k_\gamma \gamma) \,,
\ee
with $k_\gamma = (k_1^2+k_2^2)^{1/2}$. So, the lens contributions to $\ell_1 = \sum_{p=h,f} \ell_{1,p}$ depend on $\kvec$ only through $k_\kappa$ and $k_\gamma$:
\bea \label{eq:ell-p}
    \ell_{1,p}(\kvec;z) =& \,\frac{(1+z)^2}{H(z)} \int \td \ln M\,\frac{\td\bar n_p}{\td \ln M} \,r\,\td r \,\td\varphi \\ 
    &\qquad \times \Big[e^{ik_\kappa\kappa_p}J_0\big(k_\gamma\gamma_p\big) - 1 -ik_\kappa\kappa_p\Big] ,
\eea
where $\kappa_p = \kappa_p(M,r,\varphi)$ and $\gamma_p = \gamma_p(M,r,\varphi)$ are the convergence and shear magnitude of the lens.

\subsection{Subhalos}
\label{sec:sub}

A field halo is one point of a Poisson process. A host halo, in addition, carries its own population of subhalos, which we treat as a Poisson population conditional on the host~\cite{Neyman:1952xx}, with the subhalo mass function and spatial distribution of Ref.~\cite{Baltabay:2026nmz}. For a host of mass $M$ at redshift $z$ and position $\vect r$, the expected number of subhalos of mass $m$ at position $\vect d$, both positions measured from the line of sight, is
\be \label{eq:sub-intensity}
    \td\nu=\frac{\td N_{\rm sub}}{\td\ln m}\,\td\ln m\;
       n_{\rm sub}(|\vect r-\vect d|)\,\td^2 d\,\frac{\td\varphi}{2\pi} \,,
\ee
where $\td N_{\rm sub}/\td\ln m$ is the subhalo mass function of the host and $n_{\rm sub}$ is the projected distribution of subhalos around it, normalized to unity.

If the subhalos carry the mass $\sum_j m_j$, the mass of the smooth host is $M_s = M - \sum_j m_j$ and the jump of the composite lens is
\be \label{eq:sub-jump}
    \q_{\rm tot} = \q_h(M_s;\vect r,\varphi) + \sum_j \q_h(m_j;\vect d_j,\varphi_j) \,.
\ee
We keep the original truncation radius for the host halo and the subhalos are pseudo-elliptical NFW lenses truncated at $r_{200}(m)$. 

The parameters $\theta$ of a host now include its subhalo realization, and the integral over $\theta$ with weight $n(z,\theta)$ in Eq.~\eqref{eq:kf-k} includes the average over the possible realizations. For a function $\Phi$ of the host and of the subhalos covering the line of sight, this average runs over the Poisson-distributed number $N$ of such subhalos and over the mass and position of each, with the intensity $\td\nu$ of Eq.~\eqref{eq:sub-intensity}:
\be \label{eq:sub-avg}
    \langle \Phi \rangle_{\rm sub} = e^{-\bar\nu} \sum_{N=0}^{\infty} \frac{1}{N!} \int \prod_{j=1}^{N} \td\nu(m_j,\vect d_j,\varphi_j)\, \Phi(\{m_j,\vect d_j,\varphi_j\}) ,
\ee
where $\bar\nu = \int\td\nu$ is the expected number of subhalos covering the line of sight. The halo contribution then becomes 
\bea \label{eq:ell-hs}
    &\ell_{1,h}(k_\kappa,k_\gamma;z) = \frac{(1+z)^2}{H(z)} \int\!\td\ln M\,\frac{\td\bar n_h}{\td\ln M} r\,\td r \td\varphi \\
    &\quad \times\Big[\big\langle e^{ik_\kappa\kappa_{\rm tot}} J_0(k_\gamma\gamma_{\rm tot})\big\rangle_{\rm sub} - 1 -ik_\kappa\big\langle\kappa_{\rm tot}\big\rangle_{\rm sub}\Big] \,.
\eea
The mean convergence splits as $\langle\kappa_{\rm tot}\rangle_{\rm sub} = \langle\kappa_h(M_s;\vect r,\varphi)\rangle_{\rm sub} + \int\td\nu\,\kappa_h(m;\vect d,\varphi)$: the subhalo sum is linear and averages to a single integral, while the host term keeps the full average because $M_s$ depends on all the subhalo masses.

\subsection{Clustering}
\label{sec:clustering}

Correlations between the lenses add the connected terms of Eq.~\eqref{eq:psi-series} to $\Psi_1$, and we keep the two-point term. The cluster expansion is built from the factors
\bea \label{eq:F}
    &F_h(\kvec;z,\theta) = \big\langle e^{i\kvec\cdot\q_{\rm tot}}\big\rangle_{\rm sub}-1 \,, \\
    &F_f(\kvec;z,\theta) = e^{i\kvec\cdot\q_f}-1 \,.
\eea
Summing over pairs of lenses,
\bea \label{eq:psi2}
    \Psi_2(\kvec)={}&\frac12\sum_{p,p'}\int\!\td z\,\td\theta\,\td z'\,\td\theta'\;
       n_p n_{p'}\,\xi_{pp'} \\
    &\times F_p(\kvec;z,\theta)\,F_{p'}(\kvec;z',\theta') \,,
\eea
where $\xi_{pp'}$ is the correlation function of the two lenses. Unlike in $\ell_1$, the factors $F_p$ carry no compensating term: the mean $\langle\X\rangle$ is set by the lens density alone and is unchanged by clustering, and $\Psi_2$, being quadratic in the $F_p$, is $O(\kvec^2)$ and first enters the variance, where it is the two-halo term.

We use linear bias,
\be \label{eq:xi}
    \xi_{pp'}=b_p b_{p'}\,D(z)D(z')\,
       \xi_{\rm lin}(|\vect x-\vect x'|) \,,
\ee
where $\vect x$ and $\vect x'$ are the comoving lens positions, $\xi_{\rm lin}$ is the linear matter correlation function today, $D$ is the growth factor with $D(0)=1$, and $b_p$ is the peak-background-split bias of the corresponding mass function. Subhalos carry the bias of their host. This replaces the one-dimensional lognormal modulation of the lens counts used in Ref.~\cite{Baltabay:2026nmz}.

In the flat-sky Limber approximation, which keeps only lens pairs at the same redshift, $\Psi_2(\kvec)=\int_0^{z_s}\td z\,\ell_2(\kvec;z)$ with
\bea \label{eq:ell2}
    \ell_2(\kvec;z)={}&\frac{1}{2H(z)}
       \int\!\frac{\td^2l}{(2\pi)^2}\,P_{\rm lin}(l,z) \\
       &\times B(\kvec,\vect l;z)\,B(\kvec,-\vect l;z) \,,
\eea
\bea \label{eq:beta}
    B(\kvec,\vect l;z)={}&H(z)\sum_{p=h,f}\int\!\td\theta\,n_p\,b_p \\
      &\times e^{i\vect l\cdot(1+z)\vect r}\,F_p(\kvec;z,\theta) \,.
\eea
Here $\vect l$ is a transverse comoving wave vector, $P_{\rm lin}(l,z)=D(z)^2P_{\rm lin}(l,0)$ is the linear matter power spectrum, $(1+z)\vect r$ is the comoving transverse position of the lens, and $H(z)$ converts the rate per unit redshift into a rate per unit comoving distance. Because the halos and filaments are truncated, $B$ stays finite as $\vect l\to0$. Dropping the transverse phase, $\vect l\cdot(1+z)\vect r\to0$, and the shear, $k_\gamma=0$, reduces Eq.~\eqref{eq:ell2} to the two-halo term of Refs.~\cite{Thiele:2018jdl,Thiele:2020rig}.

\begin{figure*}
    \centering
    \includegraphics[width=\linewidth]{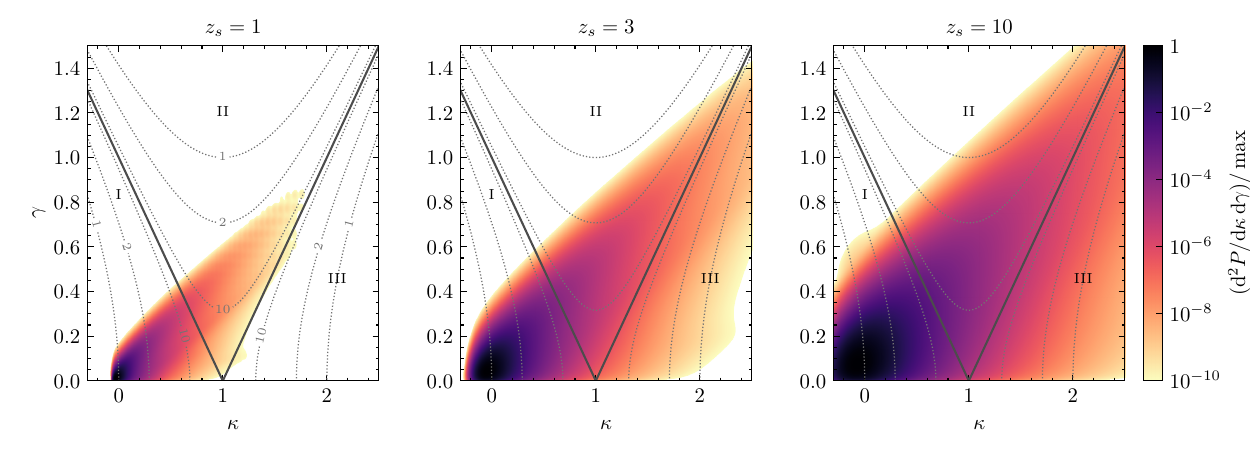}
    \vspace{-20pt}
    \caption{Joint PDF of the convergence and the shear magnitude, normalized to its maximum, at $z_s=1$, 3, and 10 for the Planck cosmology. The critical lines $\gamma = |1-\kappa|$, where $\det A=0$, are shown as solid lines and separate the regions of type I, II, and III images. Dotted lines are isocontours of magnification $\mu=1$, 2, and 10.}
    \label{fig:joint}
\end{figure*}

% ============================================================================
\section{Results}
\label{sec:pmu}
% ============================================================================

We now evaluate the characteristic function of Sec.~\ref{sec:formalism} for the lens population of Sec.~\ref{sec:population} and invert it to obtain the lensing PDFs.\footnote{The code is available at \url{https://github.com/baltabaygal/sGL_analytic}.}

\subsection{Joint distribution of convergence and shear}

The inverse Fourier transform of $\Ptil$ gives the PDF of $\X$, $P(\X) = (2\pi)^{-3} \int\!\td k_\kappa\,\td^2k_\gamma e^{-ik_\kappa\kappa - i\vect k_\gamma\cdot\vect\gamma} \Ptil(k_\kappa,k_\gamma)$. Since $\Ptil$ depends on $\vect k_\gamma$ only through its modulus, the angular integral in the shear plane gives $2\pi J_0(k_\gamma\gamma)$, and the PDF depends on the shear only through its magnitude. Multiplying by the shear-plane measure $2\pi\gamma$, the joint PDF of $\kappa$ and $\gamma$ is a Fourier transform in $k_\kappa$ and a Hankel transform in $k_\gamma$,
\be \label{eq:p2}
    \frac{\td^2 P}{\td\kappa\,\td\gamma} = \frac{\gamma}{2\pi}\int_{-\infty}^{\infty}\!\! \td k_\kappa e^{-ik_\kappa\kappa} \int_0^\infty \!\! \td k_\gamma \,k_\gamma J_0(k_\gamma\gamma)
      \Ptil(k_\kappa,k_\gamma) .
\ee

Figure~\ref{fig:joint} shows the joint PDF of $\kappa$ and $\gamma$ at $z_s=1$, 3, and 10. The PDF shows a positive correlation between $\kappa$ and $\gamma$. This is expected, as both are sourced by the same projected matter distribution. Lines of sight passing near massive halos tend to exhibit large values of both convergence and shear. Every point of the $(\kappa,\gamma)$ plane corresponds to one image type, set by the signs of $1-\kappa$ and of the determinant of the Jacobian of the lens mapping $\det A=(1-\kappa)^2-\gamma^2$~\cite{Blandford:1986zz}. Type I images are minima of the arrival-time surface ($\det A>0$, $\kappa<1$), type II are saddle points ($\det A<0$), and type III are maxima ($\det A>0$, $\kappa>1$).

\subsection{Magnification distribution} 

The image-plane magnification PDF follows from Eq.~\eqref{eq:p2} and $\mu = 1/|\!\det A|$,
\bea \label{eq:dPI}
    &\frac{\td P_I}{\td\mu} = \int\!\td\kappa\,\td\gamma\,\frac{\td^2 P}{\td\kappa\,\td\gamma}\,\delta\Big(\mu - \frac{1}{|(1-\kappa)^2-\gamma^2|}\Big) \\[4pt]
    &= \frac{1}{2\mu^2}\bigg[\int_{\mathcal{C}_-}\frac{\td\kappa}{\gamma_-} \frac{\td^2 P}{\td\kappa\,\td\gamma}\Big|_{\gamma_-} + \int_{-\infty}^{\infty} \frac{\td\kappa}{\gamma_+}\frac{\td^2 P}{\td\kappa\,\td\gamma}\Big|_{\gamma_+}\bigg] ,
\eea
where $\gamma_\pm=\sqrt{(1-\kappa)^2\pm\mu^{-1}}$ and $\mathcal{C}_- = \{\kappa: (1-\kappa)^2\ge \mu^{-1}\}$. The first integral collects the positive-parity images, minima for $\kappa<1-\mu^{-1/2}$ and maxima for $\kappa>1+\mu^{-1/2}$, and the second the saddle points. Weighting by the area factor $\mu^{-1}$ gives the source-plane PDF
\be \label{eq:PS}
    \frac{\td P_S}{\td\mu} = \mu^{-1} \frac{\td P_I}{\td\mu} \,.
\ee

\begin{figure*}
    \centering
    \includegraphics[width=\linewidth]{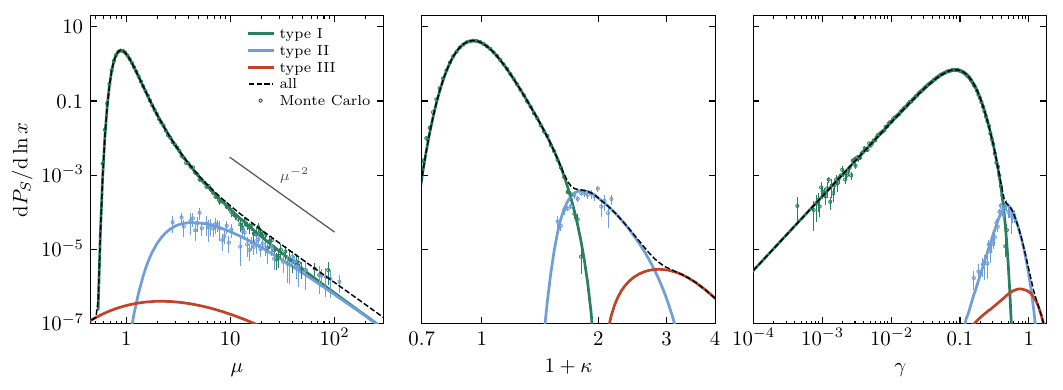}
    \vspace{-20pt} 
    \caption{PDFs of the magnification $\mu$ (left), the convergence $1+\kappa$ (middle), and the shear magnitude $\gamma$ (right) in the source plane at $z_s=5$ for the Planck cosmology, with $x$ denoting the variable on the horizontal axis. The solid curves show the contributions of type I (minima), II (saddle points), and III (maxima) images computed with the jump-process formalism, the dashed curve their sum, and the dots the Monte Carlo of Ref.~\cite{Baltabay:2026nmz}.}
    \label{fig:image-types}
\end{figure*}

Figure~\ref{fig:image-types} shows separately the contributions from the three image types to the PDFs of $\mu$, $1+\kappa$, and $\gamma$ at $z_s=5$. The full PDFs are dominated by type I around their maxima, while the high-$\kappa$ and high-$\gamma$ tails pass to type II and, beyond, to type III. In the high-$\mu$ tail the type I and II contributions coincide, as expected for the pair of images that merge at a fold caustic, one minimum and one saddle of equal magnification. Their common slope, ${\rm d}P_S/{\rm d}\ln\mu \propto \mu^{-2}$, is the fold scaling~\cite{Blandford:1986zz}. The Monte Carlo points follow the type I and II curves within their scatter. The type III contribution lies below the reach of its $10^6$ Monte Carlo realizations and is obtained with the jump-process formalism at no extra cost.

The jump-process formalism is based on the Born approximation, in which the convergence and shear are summed along the unperturbed line of sight, and the lens equation is never solved. The computed magnification is the magnification of the image at a given sky position, and the images of a multiply imaged source enter $\td P_I/\td\mu$ as separate entries. In the high-magnification tail a single lens dominates, while the other lenses add only a small external convergence and shear, and the deflection by the dominant lens before the light reaches the others is neglected~\cite{Hilbert:2008kb,McCully:2013fga}. The $\mu^{-2}$ fold scaling is generic for smooth lenses~\cite{Blandford:1986zz} and determines the shape of the tail, but not its normalization, which depends on the abundance and inner profiles of the lenses. Linking multiple images of the same source requires solving the lens equation for the dominant lens, with the other lenses contributing correlated convergence and shear to each image. We leave this extension for future work.

\subsection{Emulator}

The \texttt{FLUMEN} emulator was trained on Monte Carlo realizations~\cite{Baltabay:2026nmz}. The jump-process formalism provides an alternative source of training data, free of sampling noise also in the high-magnification tail. We therefore extend \texttt{FLUMEN} with a second network, trained on the jump-process PDFs, and users can choose between the two networks.\footnote{The new trained model is available at \url{https://github.com/baltabaygal/flumen}.} This matters most in the high-magnification tail, where the Monte Carlo has few rays per bin and the original \texttt{FLUMEN} falls below it. 

Since the jump-process formalism generates the PDFs much faster, it also allows us to cover a wider range of cosmologies. We extend the \texttt{FLUMEN} parameter space to include spatial curvature, $\Omega_K$, and a time-dependent dark-energy equation of state, $w(a)=w_0+w_a(1-a)$. Furthermore, since the training PDFs now contain no sampling noise, we also add to the loss function a term that penalizes deviations in the first derivative of the emulated PDFs~\cite{2017arXiv170604859C}.

\begin{table}
\centering
\begin{tabular}{lcc}
\hline\hline
Parameter         & Range                       & Fiducial \\
\hline
$h$               & $[0.55,\, 0.85]$            & $0.674$  \\
$\Omega_m$        & $[0.15,\, 0.50]$            & $0.315$  \\
$\sigma_8$        & $[0.40,\, 1.50]$            & $0.811$  \\
$\Omega_b$        & $[0.03,\, 0.07]$            & $0.0493$ \\
$n_s$             & $[0.94,\, 0.99]$            & $0.965$  \\
$z_{\rm eq}$      & $[3300,\, 3500]$            & $3402$   \\
$\Omega_K$        & $[-0.15,\, 0.05]$           & $0$      \\
$w_0$             & $[-1.5,\, 0.5]$             & $-1$     \\
$w_a$             & $[-2,\, 1]: w_0+w_a < 0$    & $0$      \\
\hline
$z_s$             & $[0.2,\, 12]$               & $-$      \\
\hline\hline
\end{tabular}
    \caption{Ranges of the source redshift and the cosmological parameters of the \texttt{FLUMEN} training set, and the Planck cosmological parameters for flat $\Lambda$CDM~\cite{Planck:2018vyg} used in the figures.}
    \label{tab:param}
\end{table}

\begin{figure}
    \centering
    \includegraphics[width=\linewidth]{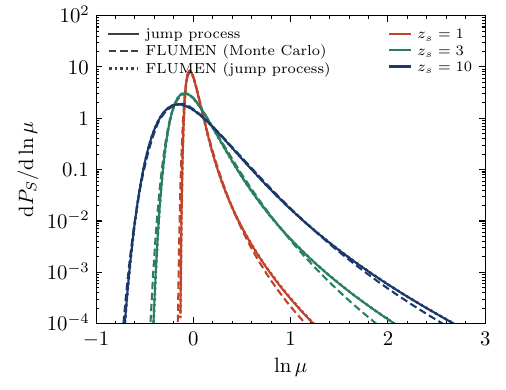}
    \vspace{-20pt} 
    \caption{Magnification PDF at $z_s = 1$, 3, and 10 for the Planck cosmology: jump-process PDFs (solid), \texttt{FLUMEN} trained on the jump-process PDFs (dotted), and the original \texttt{FLUMEN} of~\cite{Baltabay:2026nmz} trained on the Monte Carlo results (dashed).}
    \label{fig:comparison-pdf}
\end{figure}

We compute 4372 source-plane PDFs on a Latin hypercube in the ten parameters of Table~\ref{tab:param}. The source redshift spans $z_s = 0.2-12$. At lower redshift the PDF becomes so narrow, with a sharp lower edge, that the inversion requires large wave numbers on fine grids, while the lensing scatter of the observed distances falls below that from peculiar velocities~\cite{Kocsis:2005vv}. As in Ref.~\cite{Baltabay:2026nmz}, the configurations are split into 2197 for training, 1087 for validation, and 1088 for testing. On the 1088 test configurations, the maximum deviation from the master equation, as a fraction of the peak, has a median of 0.29\% and a 95th percentile of 1.7\%. The median Kullback-Leibler divergence is $10^{-6}$.

Figure~\ref{fig:comparison-pdf} compares the magnification PDFs at $z_s = 1$, 3, and 10 computed with the jump-process formalism with the two \texttt{FLUMEN} networks. The network trained on the jump-process PDFs follows them accurately over the full range of $\ln\mu$, including the high-magnification tail. The network trained on the Monte Carlo agrees with but falls below them in the high-$\mu$ tail. This reflects the sparse sampling of the tail in its training data.

\subsection{Sensitivity to $\sigma_8$}

The amplitude of matter fluctuations, $\sigma_8$, controls the abundance and clustering of the lenses. The lensing scatter of standard candles and sirens therefore carries information on $\sigma_8$~\cite{Vaskonen:2026ubd}. We study which parts of the PDF carry this information and how measurement errors degrade it. For simplicity, we vary only $\sigma_8$, with the other cosmological parameters fixed at their Planck~2018 values, and quantify the constraining power with the Fisher information. We leave a full likelihood analysis of simulated or real data, with all cosmological parameters varied jointly, for future work.

\begin{figure}
    \centering
    \includegraphics[width=\linewidth]{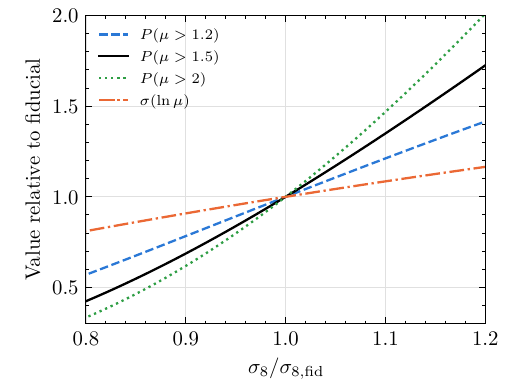}
    \vspace{-20pt} 
    \caption{Standard deviation of $\ln\mu$ and probability of a magnification above $\mu_c=1.2$, 1.5, and 2 at $z_s=1$, relative to their values at the fiducial $\sigma_8$ of the Planck cosmology, as functions of $\sigma_8$.}
    \label{fig:scaling}
\end{figure}

The tail of the magnification PDF depends on $\sigma_8$ much more strongly than its width $\sigma_{\ln \mu}$.\footnote{The fold-caustic tail ${\rm d}P_S/{\rm d}\mu \propto \mu^{-3}$ makes the variance of $\mu$ diverge logarithmically, whereas all moments of $\ln\mu$ are finite.} Figure~\ref{fig:scaling} shows this at $z_s=1$ for the standard deviation of $\ln\mu$ and the probabilities $P(\mu>\mu_c)$ of exceeding $\mu_c=1.2$, 1.5, and 2. All four increase monotonically with $\sigma_8$, and the dependence becomes stronger with increasing threshold. A 10\% change in $\sigma_8$ changes the standard deviation of $\ln\mu$ by about 9\%, but $P(\mu>1.5)$ by $-29\%$ and $+35\%$, from $P_{\rm fid}(\mu>1.5) \approx 2.2\times10^{-3}$.

A distance measurement has a finite precision, which reduces the usable sensitivity. Lensing changes the observed distances $D_L$ by a factor $\mu^{-1/2}$. Assuming a Gaussian measurement error of standard deviation $\epsilon$ in $\ln D_L$, i.e.\ a fractional distance uncertainty $\epsilon$, the logarithm of the measured magnification, $\ln \mu_{\rm obs}$, has an error of $2\epsilon$. Its probability distribution is then the true magnification PDF convolved with a Gaussian of width $2\epsilon$ in $\ln\mu$,
\be \label{eq:Peps}
    P_\epsilon(\ln\mu_{\rm obs}|\sigma_8) = \int\!\td\mu\, \frac{\td P_S(\mu|\sigma_8)}{\td\mu}\, \frac{e^{-\ln^2(\mu_{\rm obs}/\mu)/8\epsilon^2}}{\sqrt{8\pi}\,\epsilon}\,.
\ee

\begin{figure}
    \centering
    \includegraphics[width=\linewidth]{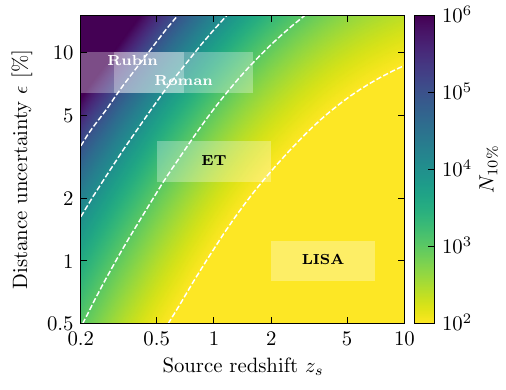}
    \vspace{-20pt} 
    \caption{Number of sources $N_{10\%}$ needed for a 10\% measurement of $\sigma_8$ with the other cosmological parameters fixed, as a function of source redshift and distance uncertainty $\epsilon$. The shaded regions indicate the approximate ranges of Rubin, Roman, ET, and LISA.}
    \label{fig:n10}
\end{figure}

The information on $\ln\sigma_8$ carried by one measured source is quantified by the Fisher information
\be \label{eq:fisher}
    F_1 = \int\!\td\ln\mu_{\rm obs}\,P_\epsilon \left(\frac{\partial\ln P_\epsilon}{\partial\ln\sigma_8}\right)^2 .
\ee
Magnifications contribute in proportion to how often they occur and to the square of how strongly the density responds to $\sigma_8$ there. For $N$ independent sources the information is $NF_1$, and by the Cram\'er-Rao bound~\cite{Tegmark:1996bz} $\sigma_8$ cannot be measured to better than $\delta\ln\sigma_8=(NF_1)^{-1/2}$. A 10\% measurement therefore requires at least $N_{10\%} = 100/F_1$ sources. Figure~\ref{fig:n10} shows $N_{10\%}$ as a function of $z_s$ and $\epsilon$, computed with the jump-process emulator of the previous subsection. It decreases with source redshift, where the lensing scatter is larger, and increases with $\epsilon$ as the noise erases the sensitive part of the PDF. It ranges from about 30 sources at $z_s=10$ and $\epsilon=0.5\%$ to around $10^5$ at $z_s=0.5$ and $\epsilon=10\%$.

To illustrate what this means for upcoming surveys, we consider benchmark samples of $10^4$ supernovae with the Roman Space Telescope at $z_s = 0.3 - 1.6$~\cite{Rose:2021nzt,Kessler:2025eib} and $10^5$ with the Rubin Observatory at $z_s = 0.2 - 0.7$~\cite{LSSTDarkEnergyScience:2018jkl,LSSTDarkEnergyScience:2021ryz}, both with $\epsilon=8\%$, and of 300 standard sirens with the Einstein Telescope (ET) at $z_s = 0.5 - 2$~\cite{Iacovelli:2022bbs,Loffredo:2024gmx} and 12 with LISA at $z_s = 2 - 7$~\cite{Tamanini:2016zlh,Mangiagli:2022niy,Mangiagli:2023ize}, with $\epsilon=3\%$ and 1\%, respectively. For simplicity, we distribute the sources uniformly in redshift and take $\epsilon$ independent of redshift. With the other cosmological parameters, the halo concentrations, and the measurement model fixed, the Fisher bound gives $\sigma_8$ to 4.8\% with Roman, 5.1\% with Rubin, 8.5\% with ET, and 19\% with LISA. These numbers are optimistic. For ET and LISA they are tighter than the 10\% and 30\% found for the same sample sizes in Ref.~\cite{Vaskonen:2026ubd}, which also varied $h$ and $\Omega_m$ in a likelihood analysis. They also depend on the assumed source populations.

% ============================================================================
\section{Conclusions}
\label{sec:conclusions}
% ============================================================================

We have shown that lensing by a population of lenses is a jump process, and that the PDF of its outcome follows from a compensated Kolmogorov-Feller equation that we solve in closed form for the characteristic function. In the Born approximation the formalism is exact for independent lenses, and the clustering of lenses enters as a hierarchy of connected correlations, of which we keep the two-point term. The formalism contains no model of the lenses. It takes their rates and jumps as input, and the lensing PDFs are then fixed, without sampling and without an adjustable parameter. Everything that remains to be modelled is therefore the lens population. We have applied the formalism to a lens population of truncated pseudo-elliptical NFW halos with subhalos and truncated cylindrical filaments, clustered through a linear-bias two-point term.

The output of the jump-process formalism is the joint PDF of the convergence and the shear, obtained by a Fourier-Hankel inversion of the characteristic function, and any observable that is a function of these two fields can be obtained from it. We have used the output for the magnification PDF and its split by image type. The PDFs are free of shot noise and respond smoothly to the cosmological parameters, which enter through the background expansion, the lens mass function and lens profiles, and their clustering. The high-magnification tail and the type III images, which the Monte Carlo cannot reach with $10^6$ realizations, come at no extra cost, and derivatives with respect to the model parameters, such as those in the Fisher information, are free of noise. This has allowed us to train \texttt{FLUMEN} on noise-free PDFs and improve it by including the derivative of the magnification PDF in the loss function. 

As a first application, we have shown that the high-magnification tail depends on $\sigma_8$ much more strongly than the width of the PDF, so that standard candles and sirens with a modest distance precision can constrain it. With the other cosmological parameters fixed, we have estimated that $10^4$ supernovae with Roman, $10^5$ with Rubin, 300 standard sirens with ET, and 12 with LISA could measure $\sigma_8$ to 4.8\%, 5.1\%, 8.5\%, and 19\%, respectively. These idealized estimates illustrate the information encoded in the magnification PDF and motivate a full analysis in which all cosmological and lens-model parameters are varied. The jump-process formalism computes the PDF, rather than sampling it, for any cosmology and lens model, and the \texttt{FLUMEN} emulator trained on these PDFs makes likelihood-based inference over the cosmological parameters practical.

Several extensions follow naturally. Training \texttt{FLUMEN} also on lens-model parameters, such as the amplitude of the halo concentrations, would allow these to be marginalized over in the likelihood. The lens model itself can be improved by including baryons through stellar cores and gas, and by replacing the pseudo-elliptical halos with projected triaxial halos. It can also be extended to dark matter models beyond cold dark matter, such as warm or fuzzy dark matter, which modify the halo abundance and profiles, and primordial black holes, which enter as an additional population of point-mass lenses. Finally, the formalism extends to two nearby sightlines, where each lens contributes a pair of jumps, so that the joint PDF of the two sources and their correlations again follow from a single integral over the lens population. Combined with the solution of the lens equation of the dominant lens, this would also describe multiply imaged sources, with the other lenses adding correlated convergence and shear to each image.

\section*{Acknowledgments} 
This work was supported by the Estonian Research Council grants TARISTU24-TK3 and TARISTU24-TK10, and the Center of Excellence program TK202 of the Estonian Ministry of Education and Research.

\bibliography{refs}

\end{document}